\documentclass[a4paper,12pt]{article}
\usepackage{amssymb,amsmath,amsthm,mathrsfs,mathtools}
\usepackage{caption}
\usepackage{cite}
\usepackage{hyperref}
\usepackage[affil-it]{authblk}
\usepackage{tikz,pgfplots,tkz-euclide}
\usetikzlibrary{arrows,positioning,calc,fadings,decorations.pathreplacing,decorations}
\def\R{\mathbf{R}}
\def\H{\mathbf{H}}
\def\q{\mathbf{q}}
\newcommand\G[1]{\Gamma(#1)}
\newcommand\eq[1]{(\ref{#1})}
\def\pa{\partial}

\def\m{{\boldsymbol\mu}}
\def\bxi{{\boldsymbol\xi}}

\newcommand\inmu[3]{I^{{#1}}_{{#2}}({#3})}

\newcommand\rt{\longrightarrow}
\def\A{\mathcal{A}}

\def\Q{{\boldsymbol Q}}

\def\tr{\text{Tr\ }}

\theoremstyle{definition}
\newtheorem{example}{Example}
\def\appell(#1,#2,#3,#4,#5,#6){F_4\big(#1,#2,#3,#4;\ #5,#6\big)}
\title{Conformal Integrals in four dimensions}
\author{Aritra Pal\thanks{email: intap@iacs.res.in}~} 
\author{Koushik Ray\thanks{email: koushik@iacs.res.in}}
\affil{Indian Association for the Cultivation of Science,\authorcr
Calcutta 700 032. India.}
\date{}
\begin{document}
\maketitle
\begin{abstract}
\noindent
We obtain analytic expressions of four-dimensional Euclidean $N$-point 
conformal integrals for arbitrary $N$ by solving a Lauricella-like system of
differential equations derived earlier. 
We demonstrate their relation to the GKZ A-hypergeometric 
systems. The conformal integrals are solutions to these expressed in terms of
leg factors and infinite series in the conformal invariant cross ratios.
\end{abstract}
\setcounter{page}{0}
\thispagestyle{empty}
\clearpage
\noindent Conformal integrals \cite{Sym} are the \emph{sine qua non} of 
theories dealing with conformal symmetry. 
The integrals make an appearance in the
evaluation of Feynman diagrams in quantum field theories
\cite{libine,BMK}, as well as studying renormalization groups \cite{WK}. 
In particular, they furnish representations of the conformal group. Conformal
blocks, which in turn determine the correlation functions of a conformal
field theory, are expressed in terms of conformal integrals.
The integrals for $N$ points have been evaluated in certain cases,
for relatively small values of $N$, looked at from different angles 
and various methods have been employed to this end 
\cite{rosenhaus,parikh1,parikh2,parikh3,gaudin1,gaudin2,gaudin3,gpz,AB,PP,
fs1,fs2,do1,mack}.
We present a general method to obtain analytic expressions of
four-dimensional $N$-point conformal integrals as infinite series in 
terms of conformal invariants, namely, the cross ratios, 
obtained as solutions to previously derived Lauricella-like equations 
\cite{PR}. We derive explicit expressions  of  the conformal integrals 
for arbitrary $N$, by showing that the Lauricella-like equations are solved
by certain GKZ A-hypergeometric functions.

Let us outline the strategy before presenting the details of the
computation.  
We restrict to the four-dimensional Euclidean space $\R^4$, 
indicating generalisation to higher dimensions at the end.
As a normed vector space $\R^4$ can be identified with the space of
quaternions $\H$, the norm-squared  being equal to the determinant of 
a quaternion.  The conformal group of $\R^4$  
is the M\"obius group
of $2\times 2$ block matrices, each block being a quaternion \eq{sl2h:def}.
The $N$-point conformal integral is defined in terms of
quaternions in equation \eq{INQ} to utilize this connection.
Differentiating with respect to quaternions within the integral sign, a
system of linear second-order 
differential equations \eq{lau:Q} is then obtained of which the integral is a
solution, analogous to its  two-dimensional counterpart
\cite{looi}. Next, the conformal integral
is  interpreted as the sheaf of germs of 
functions on the Fulton-MacPherson completion of the 
configuration space of ordered $N$-tuple of  points
on the Euclidean space, allowing it to be envisaged as a function of the
determinant of pairwise differences of the $N$ quaternions  \eq{INQ0}. 
Inserting it as an ansatz in \eq{lau:Q}   
leads to a Lauricella-like 
system of differential equations \eq{eq:IQ0} for the invariant part
of the conformal integral written  in terms of the cross ratios
\cite{PR},  generalising the Lauricella system 
for  the two-dimensional case  \cite{looi}.  
In the current article we observe that this
system of equations when cast in the form \eq{LI} 
is satisfied by the solution of a 
GKZ A-hypergeometric system
\eq{gkz1} and \eq{gkz2}
corresponding to a matrix of  exponents of the norm of pairwise
differences of the
quaternions under the M\"obius transformation. 
The solutions are then explicitly obtained as infinite series
\eq{Npt}. We discuss the examples of $N=4,5,6$ at length. 
These are consistent with previously obtained results \cite{rosenhaus},
but to the best of our knowledge the general solution has not appeared in
literature before. 

We now elaborate on the procedure, starting with a recount of the
derivation of the differential equations \cite{PR}. The conformal or M\"obius
group of $\R^4\cup\{\infty\}$ is isomorphic to a certain group of
 matrices written as $2\times 2$ blocks of quaternions
\cite{wilker,porteous}, namely,
\begin{equation}
        \label{sl2h:def}
        SL(2,\H) = \left\{\left.\begin{pmatrix}A&B\\C&D\end{pmatrix}
                \right| |AC^{-1}DC-BC|=1; A,B,C,D\in\H\right\}.
\end{equation}
The M\"obius group acts on a quaternion $Q$ as 
\begin{equation}
        \label{sl2h}
        Q\longmapsto Q'=(AQ+B)(CQ+D)^{-1}.
\end{equation} 
A real Euclidean   
four-vector  $\q=(q_0,q_1,q_2,q_3)$ in $\R^4$ is fashioned into a quaternion as
\begin{equation}
        \label{Q2}
        Q=\begin{pmatrix}
                q_0+iq_3&q_1+iq_2\\-q_1+iq_2&q_0-iq_3
        \end{pmatrix}.
\end{equation}
The determinant of the quaternion $Q$ is the Euclidean norm-squared
of the four-vector, written as
\begin{gather}
        \label{norm}
         |Q|  = q_0^2+q_1^2+q_2^2+q_3^2.
\end{gather}
The determinant of the 
difference of two quaternions, denoted $Q_{ij}=Q_i-Q_j$ from now on,
transforms under the conformal transformation \eq{sl2h} as
\begin{equation}
	\label{QQp}
	|Q'_{ij}| = |CQ_i+D|^{-1}|CQ_j+D|^{-1}|Q_{ij}|.
\end{equation}
A conformal integral is defined in terms of quaternions as
\begin{equation}
        \label{INQ}
        \inmu{\m}{N}{\Q} = \int\frac{d^4Q}{
                |Q-Q_1|^{\mu_1}|Q-Q_2|^{\mu_2}\cdots|Q-Q_N|^{\mu_N}},
\end{equation}
where $\Q$ denotes an $N$-tuple of quaternions, $\Q=(Q_1,Q_2,\cdots,Q_N)$,
$\m = (\mu_1,\mu_2,\cdots,\mu_N)$ is an $N$-tuple of real numbers and 
\begin{equation}
                \label{vol4}
                d^4Q =dq_0\wedge dq_1\wedge dq_2\wedge dq_3
\end{equation}
denotes the volume form of $\R^4$, the integral being over the whole space. The integral  transforms under the M\"obius transformation \eq{sl2h}  as 
\begin{equation}
	\label{IQtrans}
\inmu{\m}{N}{\Q'} = 
|CQ_1+D|^{\mu_1}|C Q_2+D|^{\mu_2}\cdots
|CQ_N+D|^{\mu_N} 
I_N^{\m}(\Q),
\end{equation}
provided $|\m|=\mu_1+\mu_2+\cdots+\mu_N=d$. Here, $d=4$.
Representations of the M\"obius group $SL(2,\H)$ may be
constructed out of $|Q_{ij}|$ and $I^{\boldsymbol\mu}_N({\Q})$.
Our goal is to obtain the conformal integral as a solution to a system of
differential equations. The system, which generalizes the Lauricella system
appearing in two dimensions, is set up by differentiating the integral
\eq{INQ} with respect to the $Q_i$ under the integration sign. 
Let us denote the integrand of \eq{INQ} as
\begin{equation}
\label{ingnd}
F_N^\m(Q,\Q)=\prod_{i=1}^N\frac{1}{|Q-Q_i|^{\mu_i}}.
\end{equation}
Denoting the
matrix component of a quaternion $Q_i$ by $(Q_i)_{ab}$, with $a,b=1,2$, as
defined in \eq{Q2}, we have 
\begin{equation}
\label{deriF}
\frac{\partial{F_N^\m(Q,\Q)}}{\partial{(Q_i)_{ba}}}=\mu_i(Q-Q_i)^{-1}_{ab}F_N^\m(Q,\Q).
\end{equation}
Differentiating twice and using the identity
\begin{equation}
\label{iden}
(Q-Q_i)^{-1}Q_{ij}(Q-Q_j)^{-1}=(Q-Q_i)^{-1}-(Q-Q_j)^{-1}
\end{equation}
we derive 
\begin{equation}
\label{lauF}
\sum_{b,c=1}^2(Q_{ij})_{bc}\frac{\partial}{\partial{(Q_i)_{ba}}}\frac{\partial{F_N^\m(Q,\Q)}}{\partial{(Q_j)_{dc}}}
=\mu_i\mu_j\left[(Q-Q_i)^{-1}-(Q-Q_j)^{-1}\right]_{ad}F_N^\m(Q,\Q),
\end{equation}
with $i\neq j$.  
Using \eq{deriF} and \eq{lauF} 
to perform the differentiations under the integral sign in \eq{INQ} 
we arrive at the Lauricella-like equation \cite{PR},
\begin{equation}
	\label{lau:Q}
	\sum_{b,c=1}^2(Q_{ij})_{bc}\frac{\pa}{\pa (Q_i)_{ba}}
	\frac{\pa\inmu{\m}{N}{\Q}}{\pa (Q_j)_{dc}} =
	\mu_j\frac{\pa\inmu{\m}{N}{\Q}}{\pa (Q_i)_{da}}
	-\mu_i\frac{\pa\inmu{\m}{N}{\Q}}{\pa (Q_j)_{da}},
\end{equation}
where $i,j=1,2,\cdots,N$ and $i\neq j$.
 
In order to obtain explicit expressions for the conformal integrals as
a solution to \eq{lau:Q} we first
interpret the solutions as sheaf of germs of
functions on the Fulton-MacPherson completion of the 
configuration space 
        \begin{equation}
\label{conf:N}
                C_N(M)=M^N\setminus\{q_i\in M,\ q_i\neq q_j;\
i,j=1,2,\cdots,N\}
        \end{equation}
of $N$ non-coalescing
points on the Euclidean space,
$M=\R^4\cup\{\infty\}$.
The Fulton-MacPherson completion is furnished by the embedding \cite{kmv,sinha}
        \begin{equation}
\label{conf:sinha}
                \begin{split}
                       C_N(M)&\hookrightarrow M^N\times
\big(S^3\big)^{\binom{N}{2}}
                \times[0,\infty]^{\binom{N}{3}},\\
                        \big(q_1,q_2,\cdots,q_N\big)\longmapsto
&\big(q_1,q_2,\cdots,q_N,v_{12},\cdots,v_{(N-1)N},a_{123},
                        \cdots,a_{(N-2)(N-1)N}\big),
                \end{split}
        \end{equation}
where every
\begin{equation} 
\label{conf:v}
v_{ij}=\frac{Q_{ij}}{|Q_{ij}|}
\end{equation}
describes a three-sphere $S^3$ and the scalars
\begin{equation} 
\label{conf:a}
a_{ijk}=\frac{|Q_{ij}|}{|Q_{ik}|}
\end{equation}
are non-negative real numbers. 
We used the correspondence \eq{Q2} to express $v_{ij}$ and $a_{ijk}$ in terms of
quaternions.  With this interpretation 
the conformal integral can be expressed in terms of the variables 
$a_{ijk}$, while 
invariance under translation and rotation forbids a
representation to depend on $Q_i$ alone and $v_{ij}$, respectively.
However, a product of powers of $a_{ijk}$ can be uniquely written as a product of powers 
of $|Q_{ij}|$.
Comparing \eq{QQp} and \eq{IQtrans} we conclude that a product of $|Q_{ij}|$ 
with appropriate exponents reproduces the transformation property of the
conformal integral under the M\"obius group. While this takes care of the equivariant part, 
the conformal integral, in
general, is also a function of conformal invariants, for example,
\begin{equation}
\label{def:chi}
|\chi_{ijkl}|=a_{ijk}a_{lkj},
\end{equation}
obtained as the determinant of the quaternion 
$\chi_{ijkl}=Q_{ij}Q_{ik}^{-1}Q_{kl}Q_{jl}^{-1}$.
All the $|\chi|$'s  
can be expressed in terms of $N_0= N(N-3)/2$ conveniently chosen invariants.  
We  refer to these special invariants as the cross ratios from now on 
and denote by $\xi$. We often collect the cross ratios in a vector
\begin{equation}
\label{xrat}
\bxi = \big(\xi_1,\xi_2,\cdots, \xi_{N_0}\big).
\end{equation} 
The number $N_0$ is actually an upper
bound for $N\geqslant 7$ in four dimensions \cite{gaudin2}. 
The cross ratios satisfy relations amongst themselves. 
We shall ignore this subtlety here as the conformal integrals may be obtained in those
cases by restricting to the subspace of independent cross ratios using the
relations.  The cross ratios are written as
products of ratios of $|Q_{ij}|$ as
\begin{equation}
	\label{xiQ}
	\xi_A = \prod_{\substack{i,j\\1\leq i\neq j\leq N}}
	|Q_{ij}|^{\tfrac{1}{2}\alpha^A_{ij}},
\end{equation}
where $A=1,2,\cdots, N_0$, and each $\alpha^A_{ij}$ is an integer, satisfying
\begin{equation}
	\label{alpha}
	\begin{split}
		\alpha^A_{ji}=\alpha^A_{ij}, 
\quad\alpha^A_{ii}=0, \forall i;\quad
		\sum_{j=1}^N\alpha^A_{ij}=0, \forall i,
	\end{split}
\end{equation}
for each $A$. The factor of $\tfrac{1}{2}$ is accounted for by 
the symmetry of the
cross ratios under the exchange of $i$ and $j$.
From this discussion it follows that the conformal integral \eq{INQ} as a
function on the Fulton-MacPherson completion of the configuration space of $N$
points on $M$ can be expressed as 
\begin{equation}
	\label{INQ0}
	\inmu{\m}{N}{\Q} = \prod_{\substack{i,j\\1\leq i\neq j\leq N}} 
	|Q_{ij}|^{\tfrac{1}{2}\beta_{ij}} I_0(\bxi),
\end{equation}
where $I_0(\bxi)$ is a function of the cross ratios $\bxi$, and 
\begin{equation} 
\label{betasumQ}
\quad\beta_{ji}=\beta_{ij},\quad \beta_{ii}=0, \forall i;\quad
\sum_{j=1}^N\beta_{ij}=-\mu_i;\quad
\sum_{\substack{i,j\\1\leq i<j\leq N}}\beta_{ij}=-|\m|/2=-d/2,
\end{equation}
The conformal integral is, therefore, a function of the determinant $|Q_{ij}|$ of the 
quaternions.

Plugging in \eq{INQ0} as an ansatz and taking trace over the
matrix indices $a,b$ of quaternions, the differential equation \eq{lau:Q} 
gives rise to a system of equations for the invariant part, namely \cite{PR},
\begin{multline}
	\label{eq:IQ0}
		\sum_{A,B}
		\sum_{\substack{k,l\\1\leq k,l\leq N\\
		k\neq i, l\neq j}}\alpha^A_{ik}\alpha^B_{jl}
		\tau_{ijkl}\
	\xi_A\xi_B\pa_A\pa_B I_0(\bxi)\\
		+\sum_A\Bigg(4\alpha^A_{ij}
		+\sum_{\substack{k,l\\1\leq k,l\leq N\\k\neq i, l\neq j}}
		\left(\alpha^A_{ik}\alpha^A_{jl}
		+\alpha^A_{ik}\beta_{jl}
		+\alpha^A_{jl}\beta_{ik}\right)\tau_{ijkl}\Bigg)
	\xi_A\pa_AI_0(\bxi)\\
		+\Bigg(4\beta_{ij}
		+\sum_{\substack{k,l\\1\leq k,l\leq N\\k\neq i, l\neq j}}
		\beta_{ik}\beta_{jl}\tau_{ijkl}\Bigg)I_0(\bxi)=0,
\end{multline}
where $\tau_{ijkl}=\tr{\chi_{ijkl}}$ and $\pa_A=\tfrac{\pa}{\pa\xi_A}$ 
for $A,B = 1,2,\cdots, N_0$. In order to express this set of equations
in terms of cross ratios alone we need to express the trace of 
$\chi_{ijkl}$ in terms of its determinant. From the identity \cite{PR}
\begin{equation} 
	\chi_{ijkl}\chi_{ijlk}=\chi_{ijkl}+\chi_{ijlk},
\end{equation}
taking determinant and using $\det (1+M)=1+\tr{M}+\det{M}$ for $2\times 2$
matrices, we derive
\begin{equation}
	\label{taudet}
	\tau_{ijkl}=1-|\chi_{lijk}|+|\chi_{ijkl}|.
\end{equation}
Using this in \eq{eq:IQ0} we obtain, after rearrangement of terms, the system
of differential equations in the concise form
\begin{equation}
\label{LI}
L_{ij}I_0(\bxi) = 0, 
\end{equation}  
where  for $i\neq j$, the indices
$i,j=1,2,\cdots N$, and we define the differential operator 
\begin{equation}
\label{lij}
L_{ij}= \sum_{k,l} \big(|\chi_{ijkl}|-|\chi_{lijk}|\big) 
\big(\sum_A\alpha^A_{ik}\theta_A+\beta_{ik}\big)
\big(\sum_B\alpha^B_{jl}\theta_B+\beta_{jl}\big)
+4\big(\sum_A\alpha^A_{ij}\theta_A+\beta_{ij}\big)+\mu_i\mu_j,
\end{equation} 
with  $A,B = 1,2,\cdots, N_0$, in terms of the logarithmic derivatives
\begin{equation} 
\label{tA}
\theta_A = \xi_A\frac{\pa}{\pa\xi_A}.
\end{equation} 
Singular values of $|\chi|$'s are kept from appearing in the equations by
choosing to cancel them in the sums at the level of symbols and using \eq{alpha}
and \eq{betasumQ} prior to expressing $|\chi|$'s in terms of the invariants. 

Let us also note that thanks to the relations 
\begin{equation}
|\chi_{jilk}|=|\chi_{ijkl}|,\quad
|\chi_{kjil}|=|\chi_{lijk}|,
\end{equation} 
the operators $L_{ij}$ and $L_{ji}$ give rise to  identical 
equations, leaving $N(N-1)/2$
equations in \eq{LI}. Moreover, since 
the ratios $|\chi_{ijkl}|$ and $|\chi_{lijk}|$ are interchanged
under the exchange of the indices $j$ and $l$, we have 
\begin{equation}
\label{Lsum}
\sum_{j=1}^N L_{ij}=0,
\end{equation} 
for each value of $i$. This takes away another $N$ equations, so that \eq{LI} is a system of
$N_0=N(N-3)/2$ independent ones. Hence, we have $N_0$ linear second order partial 
differential equations to solve in order to obtain $I_0$ as a function of the
same number of variables, $\bxi$. We choose the ones 
from the $L_{ij}$ by discarding the $N-1$ equations coming from $L_{1i}$ for
$i=2,3,\cdots, N$ and also $L_{23}$. 

We now describe the method of solving \eq{LI}.
Let us introduce another notation for later use. Expressing the determinant $|\chi|$ defined
in \eq{def:chi} in terms of the cross ratios as
\begin{equation}
\label{chi:g}
|\chi_{ijkl}| = \prod_{A=1}^{N_0} \xi_A^{\gamma^A_{ijkl}}, 
\end{equation} 
and using \eq{xiQ}, the consistency of the definition of $|\chi|$ requires
\begin{equation}
\label{ag}
\sum_{A=1}^{N_0} \alpha^A_{ab} \gamma^A_{ijkl} = 
(\delta_{ai}\delta_{bj} + \delta_{aj}\delta_{bi} + \delta_{ak}\delta_{bl} +
\delta_{al}\delta_{bk}) - 
(\delta_{ai}\delta_{bk} + \delta_{ak}\delta_{bi} + \delta_{aj}\delta_{bl} +
\delta_{al}\delta_{bj}), 
\end{equation} 
where a $\delta$ denotes a Kronecker delta. 
This relation can be inverted using a Gram matrix to express $\gamma$'s in terms of
the $\alpha$'s. Explicitly,
\begin{equation}
\label{gg}
\sum_{A=1}^{N_0} \sum_{a,b=1}^N \alpha^B_{ab}\alpha^A_{ab} \gamma^A_{ijkl} = 
2 \big(\alpha^B_{ij}+\alpha^B_{kl}-\alpha^B_{ik}-\alpha^B_{jl}\big).
\end{equation} 
Clearly,  \eq{chi:g} fails to hold in the instances wherein 
$\chi$ is null or singular. As mentioned above, those will not appear in the formul{\ae} below. 

We now proceed to obtain solutions to the system of $N_0$ equations \eq{LI}.
First let us define the differential operators,
\begin{equation} 
\label{Lhat}
\pa_{ij}=\frac{\pa}{\pa |Q_{ij}|},\quad
\theta_{ij}= |Q_{ij}|\pa_{ij},
\end{equation} 
\begin{equation}
\begin{split} 
\widehat{L}_{ijkl} &= \pa_{ij}\pa_{kl}-\pa_{ik}\pa_{jl}\\
&=\tfrac{1}{|Q_{ij}||Q_{kl}|}
\big( 
\theta_{ij}\theta_{kl} -|\chi_{ijkl}|\theta_{ik}\theta_{jl}
\big).
\end{split}
\end{equation} 
From \eq{xiQ}, \eq{INQ0} and \eq{tA} we obtain 
\begin{equation}
\theta_{ij}\inmu{\m}{N}{\Q} = 
\prod_{\substack{m,n\\1\leq m\neq n\leq N}} 
	|Q_{mn}|^{\tfrac{1}{2}\beta_{mn}}
\big(\sum_A\alpha^A_{ij}\theta_A+\beta_{ij}\big)I_0(\bxi),
\end{equation} 
so that 
\begin{equation}
\widehat{L}_{ijkl} \inmu{\m}{N}{\Q}
=
\left(\prod\limits_{\substack{m,n\\1\leq m\neq n\leq N}} 
	|Q_{mn}|^{\tfrac{1}{2}\beta_{mn}}\right)
\frac{1}{|Q_{ij}||Q_{kl}|}
L_{ijkl} I_0(\bxi),
\end{equation} 
where
\begin{equation}
\label{L4I0}
L_{ijkl}=\big(\sum_A\alpha^A_{ij}\theta_A+\beta_{ij}\big)
\big(\sum_B\alpha^A_{kl}\theta_B+\beta_{kl}\big) 
-|\chi_{ijkl}| 
\big(\sum_A\alpha^A_{ik}\theta_A+\beta_{ik}\big)
\big(\sum_B\alpha^A_{jl}\theta_B+\beta_{jl}\big). 
\end{equation} 
Requiring the conformal integral \eq{INQ0}, which is but a function of
$|Q_{ij}|$ treated as independent variables, to satisfy 
\begin{equation}
\label{L0}
L_{ijkl} I_0(\bxi)= 0,
\end{equation} 
or, equivalently,
\begin{equation}
\label{Lhat0} 
\widehat{L}_{ijkl}\inmu{\m}{N}{\Q}=0,
\end{equation} 
 we obtain the equation
for the invariant part.

The crucial observation in the present article is that
the equation \eq{LI} is  obtained from this
by summing over the $k$ and $l$ indices as
\begin{equation}
\label{LL}
L_{ij}I_0(\bxi)=\sum_{k,l=1}^N
L_{lijk} I_0(\bxi)
-\sum_{k,l=1}^N L_{ijkl} I_0(\bxi),
\end{equation} 
where the symmetry of $Q$, $\alpha$ and $\beta$ with respect to the indices
has been used.
We have indicated the sum in the two terms separately, since it is easier to
derive \eq{LL} by performing the sums on the RHS before subtracting.
It can be verified by explicit computation that only the operators $L_{ijkl}$
with all the four indices distinct appear in the final expression $L_{ij}$.
Many of these are, in turn, related through the inter-relations 
among the $|\chi|$'s. We need to consider only a few of these operators 
in order to obtain $I_0$.

Thus, a simultaneous solution of \eq{L0} for the operators that appear in
$L_{ij}$ is a solution to \eq{LI}. For a given set of $\alpha$, the equation
\eq{L0} is solved using the Frobenius' method with 
\begin{equation}
\label{frob1}
\prod_{A=1}^{N_0}\xi_A^{\nu_A}\sum_{n_1,n_2,\cdots, n_{N_0=0}
}^{\infty}  C_{n_1,n_2,\cdots,n_{N_0}}
\xi_1^{n_1}\xi_2^{n_2}\cdots\xi_{N_0}^{n_{N_0}},
\end{equation} 
where the solutions $\nu$ to the indicial equations can be chosen in terms of
the parameters $\beta$ and the coefficients are given by the recursion
relation
\begin{equation}
\label{rec1}
\frac{C_{n_1-\gamma^1_{ijkl},n_2-\gamma^2_{ijkl},\cdots,
n_{N_0}-\gamma^{N_0}_{ijkl}}}{C_{n_1,n_2,\cdots, n_{N_0}}}=
\frac{\big(\sum_A\alpha^A_{ij}(n_A+\nu_A)+\beta_{ij}\big)
\big(\sum_B\alpha^B_{kl}(n_B+\nu_B)+\beta_{kl}\big)}{%
\big(\sum_A\alpha^A_{ik}(n_A+\nu_A)+\beta_{ik}+1\big)
\big(\sum_B\alpha^B_{jl}(n_B+\nu_B)+\beta_{jl}+1\big)},
\end{equation} 
with the $\gamma$'s obtained from \eq{gg}.
The coefficients $C_{n_1,n_2,\cdots,n_{N_0}}$ can now be written in 
terms of Gamma functions involving the
combinations appearing within the braces.

Solving the system  \eq{LI} thus reduces to the combinatorial
problem of obtaining
the exponents of $|Q|$'s in \eq{xiQ}, that is, the $\alpha$'s, 
and expressing $\nu$'s in terms of $\beta$'s. 
In order to obtain the $\alpha$'s we consider each 
$|Q_{ij}|$ in turn, which transforms according to \eq{QQp} with a 
factor for each 
of the indices $i$ and $j$. 
Let us form a matrix  from the M\"obius transformation of $|Q_{ij}|$. 
From \eq{QQp} we note that it transforms by two factors, $(CQ_i+D)$ and 
$(CQ_j+D)$, with exponents $-1$ for each.  
We define an $N\times N(N-1)/2$ matrix $\A$ from this data. 
Its columns correspond to  
$|Q_{ij}|$ and rows correspond to $Q_i$. The entry of $\A$ 
in the column of $Q_{ij}$ in both the rows $i$ and $j$ is unity. 
All other entries are taken to be zero.  
The indices of the invariants under the M\"obius transformation
constitute the kernel of $\A$.
 A choice of the basis of the kernel is taken to
define the $\alpha$'s which in turn define the cross ratios $\bxi$ from \eq{xiQ}.  The
$N_0$ cross rations are determined by the transpose of the matrix of these basis 
vectors, denoted $v$, which is an $N_0\times N(N-1)/2$
matrix. Let us exemplify this construction
with the example of $N=4$. 
In this cases, the matrix $\A$ is given by 
\begin{equation}
        \label{amat4}
\A=\bordermatrix{%
& Q_{12}& Q_{13} & Q_{14} & Q_{23} & Q_{24} & Q_{34}\cr
Q_1& 1&1&1&0&0&0\cr
Q_2& 1&0&0&1&1&0\cr
Q_3& 0&1&0&1&0&1\cr
Q_4& 0&0&1&0&1&1\cr
}.%
\end{equation}
This encodes, for example, the fact that $|Q_{12}|$ transforms by
factors involving  $Q_1$ and $Q_2$, both having exponent $-1$,  
but does not contain factors involving $Q_3$ or $Q_4$, as can be read off from \eq{QQp}.
Let us denote the entries of $\A$ by $a_{i,jk}$, with $j<k$. 
The kernel is two-dimensional. Its transpose with a certain choice of basis vectors
is
\begin{equation}
        \label{v4}
v=\bordermatrix{%
& Q_{12}& Q_{13} & Q_{14} & Q_{23} & Q_{24} & Q_{34}\cr
\xi_1& 1&0&-1&-1&0&1\cr
\xi_2&0&1&-1&-1&1&0
}.%
\end{equation}
The $A$-th row of $v$ gives  $\alpha^A$, for example, $\alpha^1_{12}=1$,
$\alpha^2_{34}=0$ etc, so that $v=(\alpha^A_{ij})$.
Let us recall that in our notation the GKZ A-hypergeometric 
system corresponding to the matrix $\A$ is given by  \cite{hly}
\begin{gather}
\label{gkz1}
\left(\sum_{\substack{j,k=1\\j<k}}^N a_{i,jk} 
\frac{\pa}{\pa|Q_{jk}|} -\mu_i\right)f=0,\quad \forall i,\\
\label{gkz2}
\prod_{\alpha^A_{ij}>0} \left(\frac{\pa}{\pa|Q_{ij}|}\right)^{\alpha^A_{ij}} f
-\prod_{\alpha^A_{ij}<0} \left(\frac{\pa}{\pa|Q_{ij}|}\right)^{-\alpha^A_{ij}}f
=0,\quad
\forall A.
\end{gather} 
The operators acting on $f$ form an ideal in the Weyl algebra corresponding to 
the matrix $\A$.
It can be checked that these imply \eq{Lhat0}. Hence, $L_{ijkl}$ belong to
the GKZ ideal. Inserting \eq{INQ0} for $f$ the first set \eq{gkz1} is satisfied
using \eq{betasumQ}. In order to obtain $I_0(\bxi)$ it thus suffices to solve
the second set of equations \eq{gkz2}.
Expressing the GKZ operators in terms of logarithmic variables, a series
solution to these equations are obtained with its coefficients satisfying
\eq{rec1}. Hence the invariant $I_0(\bxi)$ is given by the GKZ A-hypergeometric function corresponding to the matrix $\A$

We now present examples for $N=4,5,6$. The general expression can be similarly
written.
\begin{example}
\label{ex4}
For four points, $N=4$, six operators $L_{ij}$, $i,,j=1,2,3,4$, $i<j$,
are to be
considered in \eq{LI}. Two of such operators determine the rest through the
relations
\begin{equation} 
\begin{gathered}
L_{12}=L_{34},L_{13}=L_{24},L_{14}=L_{23}, \\
L_{12}+L_{13}+L_{14}=L_{23}+L_{24}+L_{34}=0.
\end{gathered}
\end{equation} 
Choosing $L_{24}$ and $L_{34}$ as the independent ones leads to the 
equations
\begin{equation}
\label{Ap}
\begin{split}
(\xi_1+\xi_2-1)\theta_1^2I_0
&+2\xi_1\theta_1\theta_2I_0
+\xi_1(2+\beta_{13})\theta_2I_0\\
&-[\xi_1(\beta_{14}+\beta_{23})
+(1-\xi_2)\beta_{13}]\theta_1I_0
+\xi_1\beta_{14}\beta_{23}I_0=0,\\
(\xi_1+\xi_2-1)\theta_2^2I_0
&+2\xi_2\theta_1\theta_2I_0+\xi_2(2+\beta_{13})\theta_1I_0\\
&-\left[\xi_2(\beta_{14}+\beta_{23})+(1-\xi_1)\beta_{13}\right]\theta_2I_0+\xi_2\beta_{14}\beta_{23}I_0=0,
\end{split}
\end{equation}
respectively, where $\xi_1$ and $\xi_2$ are defined from $v$ in \eqref{v4}
as
\begin{equation}
\label{x1x2}
\xi_1 = \frac{|Q_{12}||Q_{34}|}{|Q_{14}||Q_{23}|},
\qquad
\xi_2 = \frac{|Q_{13}||Q_{24}|}{|Q_{14}||Q_{23}|}.
\qquad
\end{equation} 
Using the freedom of choice of $\beta$'s from \eq{betasumQ} to set
$\beta_{24}$ and $\beta_{34}$ to zero these lead to the system of equations
for the Appell function $F_4$ \cite{PR}.
Here instead of solving \eq{Ap}, we solve the system of 
equations for the operators $L_{ijkl}$, as required from \eq{LL}
without any \emph{ad hoc} choice of $\beta$'s.. 
First, we write equation \eq{LL} for $L_{24}$ and $L_{34}$. The RHS of the
two equations thus obtained contain the operators
\begin{equation}
L_{2431}, L_{3241}, L_{2413}, L_{1243}, L_{3412},
L_{2341}, L_{3421}, L_{1342}
\end{equation}
in linear combinations. 
The operators $L_{ijkl}$, however, are also related among themselves 
through the relations
\begin{equation} 
\begin{gathered}
L_{1243}=-\xi_1L_{3241}=L_{3421}\\
L_{1342}=-\xi_2L_{2341}=L_{2431}\\
\xi_1L_{2413}=-\xi_2L_{3412}=\xi_1L_{2431}-{\xi_2}L_{3421},
\end{gathered}
\end{equation} 
leaving only two of them independent. We choose these as, 
\begin{equation} 
\begin{gathered}
\label{L4pt}
L_{3421}=(\theta_1+\beta_{12})(\theta_1+\beta_{34})-\xi_1(\theta_1+\theta_2-\beta_{14})(\theta_1+\theta_2-\beta_{23}),\\
L_{2431}=(\theta_2+\beta_{13})(\theta_2+\beta_{24})-\xi_2(\theta_1+\theta_2-\beta_{14})(\theta_1+\theta_2-\beta_{23}).
\end{gathered}
\end{equation} 
The solution  $I_0(\xi_1,\xi_2)$ is annihilated by each of these operators. 
In this case there are four independent solutions corresponding to the
solutions of the indicial equations ensuing from \eq{L4pt}.
For example, for the choice of 
indices, $\nu_1=-\beta_{34}$ and $\nu_2=-\beta_{24}$,
the solution is 
\begin{equation}
\label{I04}
I_0(\xi_1,\xi_2) = \xi_1^{-\beta_{34}}\xi_2^{-\beta_{24}} P_2(\m;\bxi),
\end{equation} 
where $P_2$ is an infinite series which can be identified with the Appell
series $F_4$ up to an overall constant, namely,  
\begin{equation} 
\label{solP2}
P_2(\m;\bxi)=
\tfrac{\G{1-\mu_1-\mu_2+d/2}\G{1-\mu_1-\mu_3+d/2}}{\G{\mu_4}\G{-\mu_1+d/2}}
\sum_{n_1,n_2=0}^{\infty} 
\tfrac{\xi_1^{n_1}}{n_1!}\tfrac{\xi_2^{n_2}}{n_2!}
\tfrac{%
\G{n_1+n_2+\mu_4}
\G{n_1+n_2-\mu_1+d/2}%
}{ %
\G{1+n_1-\mu_1-\mu_2+d/2}
\G{1+n_2-\mu_1-\mu_3+d/2}
},
\end{equation} 
where we have used $|\m|=d=4$ as well as \eq{betasumQ} to replace 
linear combinations 
of $\beta$'s with $\mu$'s in deriving this expression. 
With the prefactor chosen, the solution is the Appell function $F_4$.
The four solutions to the indicial equations from \eq{L4pt} corresponds to the
four solutions to the Appell equation for $F_4$, so that the 
general expression for the $N=4$
conformal integral becomes  \cite{PR}
\begin{equation}
\label{I4:final}
I^{(\m)}_4 =
C_1(\m)f_1
+C_2(\m)f_2
+C_3(\m)f_3
+C_3(\m)f_4,
\end{equation} 
where where $C$'s are constants depending on the parameters $\m$ 
and 
\begin{equation}
\label{f14}
\begin{split}
f_1 &= \scriptstyle
|Q_{34}|^{-\mu_3-\mu_4+d/2}
|Q_{24}|^{-\mu_2-\mu_4+d/2}
|Q_{14}|^{-\mu_1}
|Q_{23}|^{\mu_4-d/2}
\appell(\mu_1, -\mu_4+d/2,1-\mu_3-\mu_4+d/2,1-\mu_2-\mu_4+d/2,\xi_1,\xi_2),\\
f_2 &= \scriptstyle
|Q_{34}|^{-\mu_3-\mu_4+d/2}
|Q_{13}|^{-\mu_1-\mu_3+d/2}
|Q_{23}|^{-\mu_2}
|Q_{14}|^{\mu_3-d/2}
\appell(\mu_2, -\mu_3+d/2,1-\mu_3-\mu_4+d/2,1-\mu_1-\mu_3+d/2,\xi_1,\xi_2),\\
f_3 &= \scriptstyle
|Q_{12}|^{-\mu_1-\mu_2+d/2}
|Q_{24}|^{-\mu_2-\mu_4+d/2}
|Q_{23}|^{-\mu_3}
|Q_{14}|^{\mu_2-d/2}
\appell(\mu_3,-\mu_2+d/2,1-\mu_1-\mu_2+d/2,1-\mu_2-\mu_4+d/2,\xi_1,\xi_2),\\
f_4 &=\scriptstyle
|Q_{12}|^{-\mu_1-\mu_2+d/2}
|Q_{13}|^{-\mu_1-\mu_3+d/2}
|Q_{14}|^{-\mu_4}
|Q_{23}|^{\mu_1-d/2}
\appell(\mu_4,-\mu_1+d/2,1-\mu_1-\mu_2+d/2,1-\mu_1-\mu_3+d/2,\xi_1,\xi_2).
\end{split}
\end{equation}
The solution is 
independent of the choice of $\beta$'s in the ansatz \eq{INQ0}, 
as expected.

Our goal is to write the conformal integral in terms of a local system on the
Fulton-MacPherson completion of the
configuration space, which possesses a canonical action of the group of
permutations of the points, to be reflected in the conformal integral. 
This action is lifted to the conformal integral as the permutation of $Q_i$
and $\mu_i$ at once.
While the integral \eq{INQ}  is invariant under these
permutations, the solution \eq{I4:final}
is not. The permutation symmetry has been broken by the choice of independent
equations, namely \eq{L4pt}. It can be restored by fixing the four constants
such that they transform appropriately under permutation of $\mu_i$.
The details of the computations to fix the constants is presented
in the Appendix. The result is
\begin{equation} 
\label{c:fix}
\begin{split}
C_1(\m) &= \G{\mu_1} \G{2-\mu_4} \G{2-\mu_1-\mu_2}\G{2-\mu_1-\mu_3},\\
C_2(\m) &= \G{\mu_2} \G{2-\mu_3} \G{2-\mu_1-\mu_2}\G{2-\mu_2-\mu_4},\\
C_3(\m) &= \G{\mu_3} \G{2-\mu_2} \G{2-\mu_1-\mu_3}\G{2-\mu_3-\mu_4},\\
C_4(\m) &= \G{\mu_4} \G{2-\mu_1} \G{2-\mu_2-\mu_4}\G{2-\mu_3-\mu_4},
\end{split}
\end{equation} 
up to an overall constant independent of $\bxi$ and $\m$, chosen to be unity
here.
\end{example}
\begin{example}
\label{ex5}
For $N=5$, we have
\begin{equation}
\A=\bordermatrix{%
& Q_{12}& Q_{13} & Q_{14} & Q_{15} & Q_{23} & Q_{24} & Q_{25} & Q_{34} & Q_{35} & Q_{45}\cr
Q_1& 1&1&1&1&0&0&0&0&0&0\cr
Q_2& 1&0&0&0&1&1&1&0&0&0\cr
Q_3& 0&1&0&0&1&0&0&1&1&0\cr
Q_4& 0&0&1&0&0&1&0&1&0&1\cr
Q_5& 0&0&0&1&0&0&1&0&1&1\cr
}%
\end{equation} 
and 
\begin{equation}
v=\bordermatrix{%
& Q_{12}& Q_{13} & Q_{14} & Q_{15} & Q_{23} & Q_{24} & Q_{25} & Q_{34} & Q_{35} & Q_{45}\cr
\xi_1& 1&1&-1&-1&-1&0&0&0&0&1\cr
\xi_2& 1&0&0&-1&-1&0&0&0&1&0\cr
\xi_3& 1&0&-1&0&-1&0&0&1&0&0\cr
\xi_4& 0&1&0&-1&-1&0&1&0&0&0\cr
\xi_5& 0&1&-1&0&-1&1&0&0&0&0\cr
}.%
\end{equation} 
In \eq{LI} we take the equations corresponding  to 
$L_{24}$, $L_{25}$, $L_{34}$, $L_{35}$, $L_{45}$ as the independent ones. 
The operators that contribute to these equations are
\begin{gather}
L_{1243},
L_{1253},
L_{1342},
L_{1352},
L_{1435},
L_{1425},
L_{1524},
L_{1534},
L_{2435},
L_{2534},
L_{2415},
L_{2451}.
\end{gather}
The simultaneous solution of the equations ensuing from these is
\begin{equation}
\label{I05}
I_0(\xi_1,\xi_2,\xi_3,\xi_4,\xi_5)= 
\xi_1^{-\beta_{45}} \xi_2^{-\beta_{35}} \xi_3^{-\beta_{34}}
\xi_4^{-\beta_{25}} \xi_5^{-\beta_{24}} P_5(\m;\bxi), 
\end{equation} 
with the series $P_5$ defined as 
\begin{small}
\begin{multline}
P_5(\m;\bxi)=
\sum_{n_1,n_2,n_3,n_4,n_5=0}^{\infty} 
\frac{\xi_1^{n_1}}{n_1!} \frac{\xi_2^{n_2}}{n_2!} \frac{\xi_3^{n_3}}{n_3!} 
\frac{\xi_4^{n_4}}{n_4!} \frac{\xi_5^{n_5}}{n_5!} 
\\\times\frac{%
1}{%
\G{1+n_1+n_2+n_3{-\mu_1-\mu_2}+d/2}
\G{1+n_1+n_4+n_5{-\mu_1-\mu_3}+d/2}}
\\\times\frac{%
1
}
{%
\G{1-n_1-n_3-n_5-\mu_4}
\G{1-n_1-n_2-n_4-\mu_5}
\G{1-n_1-n_2-n_3-n_4-n_5+{\mu_1}-d/2}
}
.%
\end{multline}
\end{small}
Plugging in \eq{INQ0}, we obtain the conformal integral
\begin{equation}
\label{I5:final}
I_5^{(\mu_1,\mu_2,\mu_3,\mu_4,\mu_5)}=|Q_{12}|^{-\mu_1-\mu_2+d/2}|Q_{13}|^{-\mu_1-\mu_3+d/2}|Q_{14}|^{-\mu_4}|Q_{15}|^{-\mu_5}|Q_{23}|^{\mu_1-d/2}P_5(\m;\bxi),
\end{equation}
independent of the choice of $\beta$'s. 
\end{example}
\begin{example}
For $N=6$ we have
\begin{equation}
\let\quad\thinspace
\A=\bordermatrix{%
\small
& Q_{12}& Q_{13} & Q_{14} & Q_{15} & Q_{16} & Q_{23} & Q_{24} & Q_{25} & Q_{26} & Q_{34} & Q_{35} & Q_{36} & Q_{45} & Q_{46} & Q_{56}\cr
Q_1& 1&1&1&1&1&0&0&0&0&0&0&0&0&0&0\cr
Q_2& 1&0&0&0&0&1&1&1&1&0&0&0&0&0&0\cr
Q_3& 0&1&0&0&0&1&0&0&0&1&1&1&0&0&0\cr
Q_4& 0&0&1&0&0&0&1&0&0&1&0&0&1&1&0\cr
Q_5& 0&0&0&1&0&0&0&1&0&0&1&0&1&0&1\cr
Q_6& 0&0&0&0&1&0&0&0&1&0&0&1&0&1&1\cr
}%
\end{equation} 
and 
\begin{equation}
\let\quad\thinspace
v=\bordermatrix{%
& Q_{12}& Q_{13} & Q_{14} & Q_{15} & Q_{16} & Q_{23} & Q_{24} & Q_{25} & Q_{26} & Q_{34} & Q_{35} & Q_{36} & Q_{45} & Q_{46} & Q_{56}\cr
\xi_1& 1&1&0&-1&-1&-1&0&0&0&0&0&0&0&0&1\cr
\xi_2& 1&1&-1&0&-1&-1&0&0&0&0&0&0&0&1&0\cr
\xi_3& 1&1&-1&-1&0&-1&0&0&0&0&0&0&1&0&0\cr
\xi_4& 1&0&0&0&-1&-1&0&0&0&0&0&1&0&0&0\cr
\xi_5& 1&0&0&-1&0&-1&0&0&0&0&1&0&0&0&0\cr
\xi_6& 1&0&-1&0&0&-1&0&0&0&1&0&0&0&0&0\cr
\xi_7& 0&1&0&0&-1&-1&0&0&1&0&0&0&0&0&0\cr
\xi_8& 0&1&0&-1&0&-1&0&1&0&0&0&0&0&0&0\cr
\xi_9& 0&1&-1&0&0&-1&1&0&0&0&0&0&0&0&0\cr
}.%
\end{equation} 
Taking $L_{24},  L_{25}, \cdots, L_{56}$ as  independent we derive
\begin{equation}
\label{I06}
I_0(\xi_1,\xi_2,\cdots,\xi_9)= 
\xi_1^{-\beta_{56}} \xi_2^{-\beta_{46}} \xi_3^{-\beta_{45}}
\xi_4^{-\beta_{36}} \xi_5^{-\beta_{35}}  %
\xi_6^{-\beta_{34}} \xi_7^{-\beta_{26}}
\xi_8^{-\beta_{25}} \xi_9^{-\beta_{24}} P_9(\m;\bxi) %
\end{equation}
 as the simultaneous solution of the corresponding GKZ system. The series
\begin{small}
\begin{multline}
P_9(\m;\bxi)=\sum_{n_1,n_2,\cdots,n_9=0}^{\infty} 
\frac{\xi_1^{n_1}}{n_1!} \frac{\xi_2^{n_2}}{n_2!} \frac{\xi_3^{n_3}}{n_3!} 
\frac{\xi_4^{n_4}}{n_4!} \frac{\xi_5^{n_5}}{n_5!} 
\frac{\xi_6^{n_6}}{n_6!} \frac{\xi_7^{n_7}}{n_7!} \frac{\xi_8^{n_8}}{n_8!} 
\frac{\xi_9^{n_9}}{n_9!} \\
\times\frac{%
1
}{%
\G{1+n_1+n_2+n_3+n_4+n_5+n_6
{-\mu_1-\mu_2}+d/2}
}%
\\\times
\frac{%
1}{%
\G{1+n_1+n_2+n_3+n_7+n_8+n_9
-\mu_1-\mu_3+d/2}}
\\\times\frac{1}{%
\G{1-n_2-n_3-n_6-n_9-\mu_4}
\G{1-n_1-n_3-n_5-n_8-\mu_5}
\G{1-n_1-n_2-n_4-n_7-\mu_6}
}
\\\times\frac{1}{%
\G{1-n_1-n_2-n_3-n_4-n_5-n_6-n_7-n_8-n_9
+\mu_1-d/2}
}.
\end{multline}
\end{small}
The conformal integral is 
\begin{small}
\begin{equation}
\label{I6:final}
I_6^{(\mu_1,\mu_2,\mu_3,\mu_4,\mu_5,\mu_6)}
=|Q_{12}|^{-\mu_1-\mu_2+d/2}|Q_{13}|^{-\mu_1-\mu_3+d/2}|Q_{14}|^{-\mu_4}|Q_{15}|^{-\mu_5}|Q_{16}|^{-\mu_6}|Q_{2,3}|^{\mu_1-d/2}P_9(\m;\bxi).
\end{equation}
\end{small}
\end{example}
These examples can be generalized to any $N$, with the matrices $v$ obtained
using \texttt{Mathematica}. 
Let us point out the strategy to fix the $\nu$'s in general.
The matrix  $v$ in the basis chosen has an exchange matrix,
one with unity on the anti-diagonal entries as the only non-zero elements on
the right. 
Then, if a $\xi_A$ has a factor of $|Q_{ij}|$
coming from this part, we choose the corresponding index $\nu_A=-\beta_{ij}$.
However, as we have shown, the choice of $\beta$'s is obliterated in the
final result.
Generally, the $N$-point conformal integral is  given by 
\begin{equation}
\label{Npt}
\inmu{\m}{N}{\Q} 
=|Q_{12}|^{-\mu_1-\mu_2+d/2}|Q_{13}|^{-\mu_1-\mu_3+d/2}
|Q_{14}|^{-\mu_4}|Q_{15}|^{-\mu_5}
\cdots|Q_{1N}|^{-\mu_N}|Q_{2,3}|^{\mu_1-d/2}P_{N_0}(\m;\bxi),
\end{equation}  
where $P_{N_0}(\m;\bxi)$ is a power series in the cross ratios,
with coefficients determined by the rows of the matrix $v$ through the
combinations appearing in \eq{rec1}.

The domain of definition of the series $P_{N_0}$
and hence the choice of independent cross ratios vary 
in computing conformal correlation functions depending on the 
specific channel. The correct
germ to be chosen is dictated by monodromy projection. 
Accordingly, the expressions presented here are to be analytically 
continued to other domains of convergence of $P_{N_0}$ by
Barnes' integrals. This can be performed since the coefficients 
are expressed in terms of Gamma
functions. 
Unlike the case of four points wherein the series can be
expressed in terms of an Appell series, however, the cases with higher
number of points the series could not be identified with known functions.  
Also, let us point out that \eq{I4:final} has four terms, expressed in terms
of the  Appell series with different parameters. This stems from the special
form of the Gale matrix \eq{v4}, in which each column is repeated twice.
For $N>4$ the columns of the Gale matrices are all different, leading to a
single series appearing in the expression for $\inmu{\m}{N}{\Q}$, as in
\eq{I5:final} and \eq{I6:final}.

To conclude, we have presented a method for computing conformal integrals in 
the four-dimensional Euclidean space with explicit expressions in terms of
infinite series of cross ratios. The method is very general and 
relates conformal integrals 
to the GKZ A-hypergeometric functions by defining them over the
Fulton-MacPherson completion of the configuration space of $N$ points on the
real Euclidean space. 
In the case of $N=4$ we have presented explicit 
expressions for the conformal integral invariant under permutation of points, 
with computational details given in the Appendix. This is required for the
integral to be a ``good" function on the configuration space. Let us remark
that in a conformal field theory the permutation symmetry is broken by the
choice of radii of convergence of operator products. Thus, in using the 
integrals in such a theory, the constants need to be fixed anew, 
preserving only the required subgroup of $S_4$. For use in other contexts the
constants are to be fixed according to physical requirements.
The conformal integrals for higher points may be treated similarly with more
cumbersome formul\ae. From the scaling properties of the
expressions it appears that the same formul{\ae} will continue to hold in any
dimension, $d$, as indicated in the expressions
in anticipation. 
Finally, the appearance of the GKZ system seems to indicate  an underlying real toric variety  associated to the configuration space of points. 

\section*{Appendix: fixing the constants in Example~\ref{ex4}}
We present the details of the computations to derive the constants 
\eq{c:fix}. This is achieved by demanding invariance of
the expression \eq{I4:final} under the action of the permutation group
${S}_4$ of $\{1,2,3,4\}$ on $Q_i$ and $\mu_i$.
It suffices to consider the generators
$\sigma_{12},\sigma_{23},\sigma_{14}$ of ${S}_4$, where $\sigma_{ij}$
denotes a cycle of the group exchanging $i$ and $j$. 
A permutation of the $Q_i$ transforms the cross ratios 
\eq{x1x2} according to 
\begin{align}
\sigma_{12}&:(\xi_1,\xi_2)\rt (\tfrac{\xi_1}{\xi_2},\tfrac{1}{\xi_2})\\
\sigma_{23}&:(\xi_1,\xi_2)\rt ({\xi_2},{\xi_1})\\
\sigma_{14}&:(\xi_1,\xi_2)\rt ({\xi_2},{\xi_1}),
\end{align}
thereby transforming the Appell functions appearing in \eq{f14}. The first
four arguments depending on $\mu_i$ also change. 
Using the transformation formul{\ae} of $F_4$, namely, 
\begin{equation}
F_4(a,b;c,d;x,y) = F_4(a,b;d,c;y,x),
\end{equation} 
which follows from the definition, and
\begin{multline}
F_4(a,b;c,d;x,y) = 
\tfrac{\G{d}\G{b-a}}{\G{d-a}\G{b}} (-y)^{-a}
F_4(a,a-d+1;c,a-b+1,\tfrac{x}{y},\tfrac{1}{y})
\\+
\tfrac{\G{d}\G{a-b}}{\G{d-b}\G{a}} (-y)^{-b}
F_4(b,b-d+1;c,b-a+1,\tfrac{x}{y},\tfrac{1}{y}),
\end{multline} 
the Appell functions can be expressed back in terms of $F_4$ with
arguments $(\xi_1,\xi_2)$.
Since the functions \eq{f14} are the solutions to the four indicial equations
associated to \eq{L4pt}, these are the germs of the local system in a
neighborhood of $\bxi=0$ forming a basis. 
Hence, $I^{(\m)}_4$ in \eq{I4:final} can be expressed in terms
of the same functions \eq{f14} with new constants. In this manner the
permutations induce an action on the $C$'s.
Let us denote the action of the generators $\sigma_{ij}$ on the constants by
\begin{equation}
C_i'(\m) = \sigma_{12}C_i(\m),\quad
C_i''(\m) = \sigma_{23}C_i(\m),\quad
C_i'''(\m) = \sigma_{14}C_i(\m),
\end{equation} 
$i=1,2,3,4$. Writing the quadruple of $C$'s as a vector we obtain a matrix
representation of the permutations. For example, 
\begin{equation}
\label{CpC}
\left(\begin{smallmatrix}
C'_1(\m)\\C'_2(\m)\\C'_3(\m)\\C'_4(\m)
\end{smallmatrix}\right) 
= \Sigma_{12}(\m)
\left(\begin{smallmatrix}
C_1(\m)\\C_2(\m)\\C_3(\m)\\C_4(\m)
\end{smallmatrix}\right), 
\end{equation} 
where $\Sigma_{12}(\m)$ denotes the transformation 
matrix under $\sigma_{12}$, and
similarly for the other two generators.
The three transformation matrices are
\begin{small}
\begin{multline}
\label{Sgma}
\Sigma_{12}(\m) = \\
\begin{psmallmatrix}
\frac{(-1)^{\mu_4}\G{3-\mu_1-\mu_4}
\G{2-\mu_1-\mu_3}}{
\G{\mu_2} \
\G{1-\mu_1}} 
& \frac{(-1)^{-\mu_1}\G{3-\mu_2-\mu _3}
\G{2-\mu_1-\mu_3}}{
\G{2-\mu_3}\G{\mu_4-1}} & 0 & 0 \\
\frac{(-1)^{-\mu _2}\G{3-\mu_1-\mu_4} 
\G{2-\mu_2-\mu _4}}{
\G{\mu_3-1} \G{2-\mu_4}} 
&\frac{(-1)^{\mu_3}\G{3-\mu_2-\mu_3} 
\G{2-\mu_2-\mu_4}}{
\G{1-\mu_2} \G{\mu_1}} & 0 & 0 \\
0 & 0 & \frac{(-1)^{-\mu_3}\G{3-\mu_1-\mu_4} 
\G{2-\mu_1-\mu_3}}{
\G{\mu_2-1}\G{2-\mu_1}} & 
\frac{(-1)^{\mu_2}\G{3-\mu _2-\mu _3} 
\G{2-\mu_1-\mu_3}}{
\G{1-\mu _3}\G{\mu _4}} \\
0 & 0 & \frac{(-1)^{\mu_1}\G{3-\mu_1-\mu _4}
\G{2-\mu_2-\mu_4}}{
\G{\mu_3}\G{1-\mu_4}} & 
\frac{(-1)^{-\mu_4} 
\G{3-\mu_2-\mu_3} \G{2-\mu_2-\mu_4}}{
\G{2-\mu_2} \G{\mu_1-1}} 
\end{psmallmatrix},
\\
\Sigma_{23}=\begin{pmatrix}
1&0&0&0\\
0&0&1&0\\
0&1&0&0\\
0&0&0&1
\end{pmatrix},
\qquad
\Sigma_{14}=\begin{pmatrix}
0&0&0&1\\
0&1&0&0\\
0&0&1&0\\
1&0&0&0
\end{pmatrix}.
\end{multline}
\end{small} 
The arguments of
the latter two are suppressed since  they do not depend on $\m$.

To fix the constants, we first impose the condition that $\sigma_{12}$ acting
twice on $C_i(\m)$ keeps them unchanged. From \eq{CpC}, then, we conclude that
the vector of $C$'s is an eigenvector of the product matrix
$\Sigma'_{12}(\m)\Sigma_{12}(\m)$
with unit eigenvalue. Here the prime on $\Sigma_{12}$ signifies that the
arguments of $\Sigma_{12}$ are permuted by $\sigma_{12}$. 
Solving the eigenvalue problem fixes the ratios
\begin{gather}
\label{c2c1}
\frac{C_2(\m)}{C_1(\m)} =
\frac{\sin\pi\mu_1\sin\pi\mu_4}{\pi\sin\pi\left(\mu_2+\mu_4\right)}
\frac{\G{1-\mu_1} \G{\mu_2} \G{2-\mu_3} \G{\mu_4-1} }{
\G{2-\mu_1-\mu_3}\G{3-\mu_1-\mu_3}},\\
\label{c4c3}
\frac{C_4(\m)}{C_3(\m)} =
\frac{\sin\pi\mu_2\sin\pi\mu_3}{\pi\sin\pi\left(\mu_2+\mu_4\right)}
\frac{
\G{2-\mu_1} \G{\mu_2-1} \G{1-\mu_3} \G{\mu_4} }{
\G{2-\mu_1-\mu_3}\G{3-\mu_1-\mu_3}}.
\end{gather} 
According to \eq{Sgma}, $C''_1(\m)=C_1(\m)$ and  $C''_2(\m)=C_3(\m)$ under
the action of $\sigma_{23}$.  
Acting \eq{c2c1} with $\sigma_{23}$ we obtain 
\begin{equation}
\label{c3c1}
\frac{C''_2(\m)}{C''_1(\m)} =
\frac{C_3(\m)}{C_1(\m)} =
\frac{\sin\pi\mu_1\sin\pi\mu_4}{\pi\sin\pi\left(\mu_3+\mu_4\right)}
\frac{\G{1-\mu_1} \G{2-\mu_2} \G{\mu_3}\G{\mu_4-1}}{
\G{2-\mu_1-\mu_2}\G{3-\mu_1-\mu_2}}.
\end{equation} 
Using this in \eq{c4c3} we obtain 
\begin{small}
\begin{equation} 
\label{c4c1}
\frac{C_4(\m)}{C_1(\m)} = 
\frac{\sin\pi\mu_1\sin\pi\mu_4}{\sin\pi\left(\mu_2+\mu_4\right)
\sin\pi\left(\mu_3+\mu_4\right)}
\frac{\G{1-\mu_1}\G{2-\mu_1}\G{\mu_4}\G{\mu_4-1}}{
\G{2-\mu_1-\mu_3}\G{3-\mu_1-\mu_3} 
\G{2-\mu_1-\mu_2}\G{3-\mu_1-\mu_2}}
\end{equation}
\end{small} 
We have thus obtained $C_2(\m)$, $C_3(\m)$ and $C_4(\m)$ in terms of
$C_1(\m)$ in \eq{c2c1}, \eq{c3c1} and \eq{c4c1}, respectively.

While we have used the invariance of the constants under $\sigma_{12}$ acting
twice up till now, we have not used the transformation
\eq{CpC} directly. Using \eq{c2c1},
\eq{c3c1} and \eq{c4c1} in \eq{CpC} we obtain the ratios
\begin{equation} 
\begin{split}
\frac{C'_1(\m)}{C_1(\m)} &= 
\frac{\G{\mu_2}\G{2-\mu_2-\mu_3}}{\G{\mu_1}\G{2-\mu_1-\mu_3}},\\
\frac{C'_2(\m)}{C_2(\m)} &= 
\frac{\G{\mu_1}\G{2-\mu_1-\mu_4}}{\G{\mu_2}\G{2-\mu_2-\mu_4}},\\
\frac{C'_3(\m)}{C_3(\m)} &= 
\frac{\G{2-\mu_1}\G{2-\mu_2-\mu_3}}{\G{2-\mu_2}\G{2-\mu_1-\mu_3}}, \\
\frac{C'_4(\m)}{C_4(\m)} &= 
\frac{\G{2-\mu_2}\G{2-\mu_1-\mu_4}}{\G{2-\mu_1}\G{2-\mu_2-\mu_4}}.
\end{split} 
\end{equation} 
Since the primed constants are the transformed ones under $\sigma_{12}$, we
deduce 
\begin{equation} 
\begin{split}
C_1(\m) &\propto {\G{\mu_1}\G{2-\mu_1-\mu_3}},\\
C_2(\m) &\propto {\G{\mu_2}\G{2-\mu_2-\mu_4}},\\
C_3(\m) &\propto {\G{2-\mu_2}\G{2-\mu_1-\mu_3}},\\
C_4(\m) &\propto {\G{2-\mu_1}\G{2-\mu_2-\mu_4}}.
\end{split}
\end{equation}  
The constants transform under the other two generators as well.
Under the action of $\sigma_{23}$  
these expressions transform to 
\begin{equation} 
\begin{split}
C''_1(\m) &\propto {\G{\mu_1}\G{2-\mu_1-\mu_2}},\\
C''_2(\m) &\propto {\G{\mu_3}\G{2-\mu_3-\mu_4}},\\
C''_3(\m) &\propto {\G{2-\mu_3}\G{2-\mu_1-\mu_2}},\\
C''_4(\m) &\propto {\G{2-\mu_1}\G{2-\mu_3-\mu_4}}.
\end{split}
\end{equation}  
The constants $C_1$ and $C_4$ remain invariant under $\Sigma_{23}$,
 while the other two are exchanged, as is seen from \eq{Sgma}. 
Incorporating extra factors thus arising we obtain 
\begin{equation} 
\begin{split}
C_1(\m) &\propto {\G{\mu_1}\G{2-\mu_1-\mu_2}}\G{2-\mu_1-\mu_3},\\
C_2(\m) &\propto {\G{\mu_2}\G{2-\mu_1-\mu_2}}\G{2-\mu_3}\G{2-\mu_2-\mu_4},\\
C_3(\m) &\propto {\G{2-\mu_2}\G{2-\mu_1-\mu_3}} \G{\mu_3} \G{2-\mu_3-\mu_4},\\
C_4(\m) &\propto {\G{2-\mu_1}\G{2-\mu_2-\mu_4}}\G{2-\mu_3-\mu_4}.
\end{split}
\end{equation}  
Furthermore,
the constants $C_2$ and $C_3$ are invariant under $\sigma_{14}$, in accordance
with $\Sigma_{14}$ in \eq{Sgma}, while $C_1$ and $C_4$ transform to
\begin{equation} 
\begin{split}
C'''_1(\m) &\propto {\G{\mu_4}\G{2-\mu_4-\mu_2}}\G{2-\mu_4-\mu_3}\\
C'''_4(\m) &\propto {\G{2-\mu_4}\G{2-\mu_2-\mu_1}}\G{2-\mu_3-\mu_1}.
\end{split}
\end{equation}  
The matrix $\Sigma_{14}$ exchanges $C_1$ and $C_4$.
Incorporating the additional factors in these two we derive the expressions
\eq{c:fix} taking the constant of proportionality as unity.

\end{document}